\documentclass[
aps,%
final,%
notitlepage,%
twocolumn,%
nobibnotes,%
nofootinbib,%
superscriptaddress,%
noshowpacs,%
centertags,
showkeys]%
{revtex4-2}

\usepackage{graphicx}
\usepackage{dcolumn}
\usepackage{bm}

\usepackage{amsmath,amssymb}
\usepackage[english]{babel}
\usepackage{graphicx}
\graphicspath{{./}{Figures/}}
\usepackage{xcolor}
\usepackage{cmap}
\usepackage{placeins}

\usepackage{hyperref}
\usepackage{soul}      
\usepackage{lipsum}    
\usepackage{blindtext} 
\usepackage{microtype} 

\definecolor{LightCyan}{rgb}{0.88,1,1}

\def\*{$^{*}$}

\def\etal{{et~al.}}
\def\be{\begin{equation}}
\def\ee{\end{equation}}
\def\ba{\begin{aligned}}
\def\ea{\end{aligned}}

\newcommand{\PWN}{\raise-0.4ex\hbox{\scalebox{0.8}{\scriptsize$P$\kern-0.05em$W$\kern-0.2em$N$}}}
\newcommand{\s}{\raise-0.1ex\hbox{\scalebox{1.2}{\scriptsize$s$}}}
\newcommand{\sla}{\;\raise0.55ex\hbox{\scriptsize$<$\kern-0.75em\raise-1.1ex\hbox{$\sim$}}\;}
\newcommand{\sga}{\;\raise0.55ex\hbox{\scriptsize$>$\kern-0.75em\raise-1.1ex\hbox{$\sim$}}\;}
\newcommand{\ssim}{\;\raise0.3ex\hbox{\tiny$\sim$}\,}
\newcommand{\sapprox}{\;\raise0.3ex\hbox{\tiny$\approx$}\,}

\def\lsim{\;\raise0.3ex\hbox{$<$\kern-0.75em\raise-1.1ex\hbox{$\sim$}}\;}
\def\gsim{\;\raise0.3ex\hbox{$>$\kern-0.75em\raise-1.1ex\hbox{$\sim$}}\;}

\def\etal{{ et al. }}

\def\ecsb{erg cm$^{-2}$ s$^{-1}$ arcsec$^{-2}$ }
\def\ecsb2{erg cm$^{-2}$ s$^{-1}$ arcsec$^{-2}$}

\def\apj{ApJ}
\def\mnras{MNRAS}
\def\nat{Nat}

\def\aap{A\&A}                   
\def\apjl{ApJ}                   

\def\ssr{Space Sci. Rev.}

\def\physrep{Phys. Reports}

\def\prl{Phys. Rev. Lett.}
\def\jcap{JCAP}

\begin{document}

\title{On the contribution of the bow shock pulsar wind nebula PSR J0437-4715 \\ to the observed fluxes of GeV-TeV positrons and antiprotons}

\author{A. E. Petrov}%
\email{a.e.petrov@mail.ioffe.ru}
\author{A. M. Bykov}
\email{byk@astro.ioffe.ru}
\affiliation{ 
Ioffe Institute, St.~Petersburg, 194021 Russia
}%

\begin{abstract}
The orbital observatories PAMELA and AMS-02 have detected a significant excess in the cosmic ray (CR) positron flux at energies above several tens of GeV. The measured values exceed those expected in models of secondary origin of positrons due to inelastic collisions of CR nuclei with the interstellar matter. This excess may be due to the annihilation or decay of hypothetical dark matter particles or, alternatively, to the contribution of local sources of primary positrons, particularly pulsars. In contrast, the antiproton fluxes observed by AMS-02 at energies above GeV are consistent with the models of their secondary origin, taking into account the uncertainties. The ratio of the observed positron flux to the antiproton flux is virtually independent of energy in the range from 60 to 400 GeV. This behavior can be understood if the observed local spectra of positrons and antiprotons in the range of tens to hundreds of GeV are formed by the same source. The relativistic winds of pulsars inject accelerated electrons and positrons into the interstellar medium. Fast-moving pulsars form pulsar wind nebulae with bow shocks (BSPWNe), which accelerate both the freshly injected positrons and electrons of the pulsar wind and the hadrons and leptons of galactic CRs from the interstellar medium via the Fermi acceleration mechanism in colliding flows. Such a system can produce identical particle spectra regardless of the site of their injection. The nearest to the Earth millisecond pulsar PSR J0437-4715 forms a pulsar wind nebula (PWN) with a bow shock observable in optical and ultraviolet wavelengths. This BSPWN is a possible candidate for the main near-Earth ``factory'' of antiparticles along with the Geminga PWN. 
Considering PSR J0437-4715, we provide the Monte Carlo simulations of particle acceleration in its BSPWN and the analytical model of anisotropic diffusion in the local interstellar medium. We show that this pulsar's contribution can explain the observed positron flux in the range from 30 GeV to 1 TeV, and simultaneously the antiproton flux at hundreds of GeV with an almost energy-independent positron-to-antiproton flux ratio.
The model allows to reproduce the observed antiparticle fluxes if $\sim 25\%$ of the PSR J0437-4715 pulsar wind power is transferred to accelerated positrons and electrons and used to re-accelerate antiprotons.
\end{abstract}


\maketitle

\selectlanguage{english}

\section{Introduction}
\label{sec:intro}
Discovered in cosmic rays more than 90 years ago, relativistic positrons continue to be a tool for the studies of fundamental problems in physics. Sensitive detections of CR positrons and antiprotons with energies up to TeV by the modern observatories PAMELA \citep{Adriani+09} and AMS-02 \citep{Aguilar+14,Aguilar+21,Aguilar+25} have revealed the positron excess.
The measured positron flux significantly exceeds the flux of secondary positrons predicted by modern numerical models (e.g., the GALPROP model; see \citealp{Trotta+11,Vladimirov+11} and references therein). These secondary positrons  are produced in reactions of CR nuclei with interstellar matter during their diffusive propagation in the Galaxy \citep{Berezinskii_astrophys_of_CR_1990,MoskalenkoStrong98,DiesingCaprioli20,Huang+20}.  
Excess positrons must be produced by some additional source of antiparticles, such as nearby pulsars \citep[e.g.,][]{Hooper+09, Yueksel+09}, dark matter particles \citep[see, e.g.,][]{SilkSrednicki84,Bertone+05,Bergstroem+08,Evoli+21}, nearby old supernova remnants \citep{Mertsch21}, or nearby molecular clouds irradiated by CR fluxes \citep{Dogiel+87}.
Modeling of the positron and antiproton fluxes produced by inelastic collisions of
primary CRs in the Galaxy allows us to constrain possible scenarios for the origin of CR antimatter \citep{Moskalenko+02,Moskalenko+03,Vladimirov+12,DiMauro+23}.
%

The decays and annihilations of hypothetical dark matter particles can generate CR antiparticles detected by near-Earth observatories.
An argument in favor of dark matter as the source of the observed antiparticles is the nearly energy-independent ratio of the observed positron and antiproton fluxes measured by the AMS-02 observatory \citep{Aguilar+16}. The similarity of the spectra of the two types of antiparticles may mean that they are accelerated by the same objects. The magnetized relativistic winds of pulsars are proposed to consist of electrons and positrons (although some models do not exclude the existence of a proton/ion component; see, e.g., \citealp{amato_arons06,AmatoOlmi2021}), but do not contain primary antiprotons. On the other hand, annihilations and decays of dark matter particles can generate both types of antiparticles. This may lead one to the conclusion that models considering pulsars as the source of detectable antiparticle fluxes in the GeV-TeV energy range face a  problem that is absent in dark matter models.
Meanwhile, the relativistic outflows from pulsars and the nebulae that they form are capable of accelerating both leptons and hadrons. Of particular interest are the PWNe of fast-moving pulsars -- those with supersonic speeds relative to the interstellar medium -- that launch bow shocks (BS). These BSPWNe can simultaneously reaccelerate charged CR particles injected from the interstellar medium and accelerate relativistic positrons and electrons injected from the pulsar wind \citep{BSPWN_2017}. Bow shocks from other stars are also discussed as possible local CR accelerators \citep[see ][]{MalkovMoskalenko21}. Energetic CR particles, accelerated by the Fermi mechanism at the shock front, will increase their energy, forming a population of re-accelerated particles with a typical spectral energy distribution $f\left(E\right)\propto E^{-s}$, $s \sga 2$. The collision of the relativistic wind with the ambient medium launches the pulsar wind's termination shock (TS), where the pulsar wind's particles (positrons and electrons, and possibly ions) are accelerated to form a power-law spectrum \citep[e.g., ][]{Kirk+00,Ellison+16}.
A specific feature of BSPWNe is a very efficient Fermi I acceleration in the zone of colliding flows \citep{Bykov+13,BSPWN_2017,Grinaldo+19,MalkovLemoine23,Aerdker+25} between two shocks -- the TS and the BS.
In this acceleration process, particles that have undergone preliminary acceleration at one of the single shocks and have gained enough energy to overcome the thickness of the layer near the contact discontinuity between the flows ($E> E_{min}$) can repeatedly diffuse through the contact discontinuity. The multiple diffusion is ensured by scattering on inhomogeneities of the turbulent magnetic field of first one and then the other flow. This occurs as long as the energy of the particles allows their confinement in the collision zone ($E<E_{max}$). The result of such acceleration process 
is the formation of a much harder spectrum with $f\left(E\right)\propto E^{-s}$, $s \sim 1$ in the range $E_{min} < E < E_{max}$.
Injection into this acceleration mechanism is possible from both shocks. This means that both positrons and electrons of the relativistic wind, pre-accelerated at the TS, and hadrons and leptons of galactic CRs, re-accelerated by the BS, can be accelerated by the same physical mechanism in the same flow system. As a result, regardless of the injection origin, the particles acquire the same spectrum, which could explain the flat positron-to-antiproton ratio if both the positrons and antiprotons are accelerated in the same nebula.

The near-Earth pulsars have been considered as the sources of primary positrons in a number of recent studies \citep[e.g., ][]{Fang+18,TangPiran19,Bykov+19,Orusa+25}. Simulations by \citet{Bykov+19} demonstrated that the entire observed positron flux in the range from tens of GeV to 1 TeV can be explained by the contribution of the nearest pulsar PSR J0437-4715, which forms a BSPWN. In this work, we will demonstrate that this object can simultaneously be a source of re-accelerated antiprotons, allowing us to explain the observed nearly energy-independent ratio of positron and antiproton fluxes at least at energies above 100 GeV. Section \ref{sec:source} presents known facts about the particle accelerator under consideration; Section \ref{sec:diffusion} discusses the propagation of particles in the local interstellar medium; Section \ref{sec:model} presents a model for calculating antiparticle fluxes (details of which can be found in the Appendix \ref{sec:appendix}); Section \ref{sec:results} gives the results of the simulation; Section \ref{sec:discussion} provides a discussion of the obtained results; and in the final Section \ref{sec:conclusion} the conclusions are presented.

\section{PSR J0437-4715 and its BSPWN}
\label{sec:source}
The old millisecond pulsar PSR J0437-4715
forms a binary system with a white dwarf, the distance to which $d$ is known with very high accuracy due to measurements of both the parallax ($d=156.3 \pm 1.3$ pc, \citealp{Deller+08}) and the time derivative of the orbital period ($d=156.79 \pm 0.25$ pc, \citealp{Reardon+16}).
The pulsar has\footnote{assuming a stiff equation of state for the matter of a neutron star with a mass of 1.44 solar masses, allowing its moment of inertia to reach $2\times10^{45} \,\mbox{g}\,\mbox{cm}^{2}$} the spin-down power of $\dot{E} = 6 \times 10^{33}\,\mbox{erg}\,\mbox{s}^{-1}$ and moves through the interstellar medium with a proper velocity whose projection onto the plane of the sky is $104.14 \pm 0.17\,\mbox{km}\,\mbox{s}^{-1}$ \citep{Rangelov+16}, which exceeds the characteristic sonic speed values in the interstellar medium. Observations of this object in H$\alpha$ \citep[see, e.g., ][]{BrownsbergerRomani14} and in the far ultraviolet range with the Hubble Space Telescope \citep{Rangelov+16} indeed revealed the presence of a bow shock ahead of the pulsar. The Chandra X-ray observations did not detect a bow shock, but revealed a faint, extended source near the pulsar associated with its PWN (relativistic wind's plasma that has passed through the termination shock).

Monte Carlo simulations of the acceleration of positrons and electrons injected from the relativistic pulsar wind of PSR J0437-4715 -- in the collision zone where the pulsar wind interacts with the interstellar medium -- allowed \citet{BSPWN_2017} to explain the observed structure and spectra of this object's BSPWN in the far ultraviolet and X-ray ranges. The simulated fluxes of accelerated positrons and electrons leaving this accelerator were then used by \citet{Bykov+19} as a source function in the developed model of charged particle propagation from PSR J0437-4715 to Earth.

\section{Diffusion in the local interstellar medium}
\label{sec:diffusion}
The parameterization of the diffusion propagation of accelerated CR particles from the source to Earth depends on the source's location. The global (average) particle diffusion coefficient in the Galaxy is determined by the expression \citep{Strong+07}:
\begin{equation}
D\left(R\right) \sim 3 \times 10^{28} \left(\frac{R}{1 \:\mbox{GV}}\right)^a \:\mbox{cm}^2\,\mbox{s}^{-1},
\label{diff_global}
\end{equation}
with $a \sim 0.3-0.6$, where $R$ is the magnetic rigidity.
Strong fluctuations of the galactic magnetic field on scales of $L \sim $ 100 pc make global diffusion in the Galaxy isotropic.

At scales of $\sim$ 100 pc and smaller, particle diffusion becomes significantly anisotropic: transport across the local large-scale magnetic field is suppressed. At scales below $L$, assuming the Kolmogorov spectrum of interstellar turbulence ($a = 1/3$), the diffusion coefficient can be estimated as
\begin{equation}
D\left(R\right) \sim 2 \times 10^{27} \left(\frac{R}{1 \:\mbox{GV}}\right)^{1/3} \:\mbox{cm}^2\,\mbox{s}^{-1}
\label{diff_meso}
\end{equation}
\citep[see ][]{Strong+07,TangPiran19}.

In the vicinity of efficient particle accelerators, magnetic field turbulence can be amplified by current instabilities generated by the accelerated particles, which can further slow down the diffusion. \citet{Abeysekara+17} calculated the particle diffusion coefficient near the pulsars Geminga and PSR B0656\,+\,14 by modeling the emission from the observed extended teraelectronvolt halo with a size $\sim$50 pc:
\begin{equation}
D\left(R\right) \sim 10^{26} \left(\frac{R}{1 \:\mbox{GV}}\right)^{1/3} \:\mbox{cm}^2\,\mbox{s}^{-1}.
\label{diff_Abeysekara}
\end{equation}

Since the distance to the Geminga pulsar is about 250 pc, a two-zone model is applicable for the diffusion from this source: with slow propagation described by the coefficient (\ref {diff_Abeysekara}) at distances up to $\sim$ 50 pc from the pulsar and with a faster diffusion corresponding to the coefficients (\ref{diff_global})-(\ref{diff_meso}) at greater distances. Such models have been developed by a number of authors  \citep[see ][]{Fang+18,TangPiran19,Manconi+20,Osipov+20,Fang24} and made it possible to explain the observed near-Earth positron flux, provided that a significant ($40-75$\%) fraction of the pulsar spin-down power $\dot{E}$ is involved. Interestingly, there are some other explanations for the slow diffusion of particles near the Geminga pulsar, including the suppression of diffusion in the direction across the large-scale magnetic field \citep{Xia+25}.

The nearest pulsar PSR J0437-4715, which is one and a half times closer to Earth than Geminga, emits only a few times less power $\dot{E}$ and appears to be a natural candidate for the role of the main near-Earth ``positron factory''. Due to the relatively small distance of $\sim 100$ pc, \citet{Bykov+19} considered the anisotropic diffusion caused by a local large-scale interstellar magnetic field directed in accordance with the modern measurements. This made it possible to explain the entire observed positron flux from tens of GeV to 1 TeV by the contribution of PSR J0437-4715, without contradicting the data on the total lepton flux and using only 30\% of $\dot{E}$. Below, we discuss in detail a model for the production of near-Earth spectra of both positrons and antiprotons involving the contribution of PSR J0437-4715, similar to the model \citep{Bykov+19}, and demonstrate that this pulsar can simultaneously make a significant contribution to the fluxes of both these types of antiparticles.

\section{Model}
\label{sec:model}
\begin{figure}
\center{\includegraphics[width=\columnwidth]{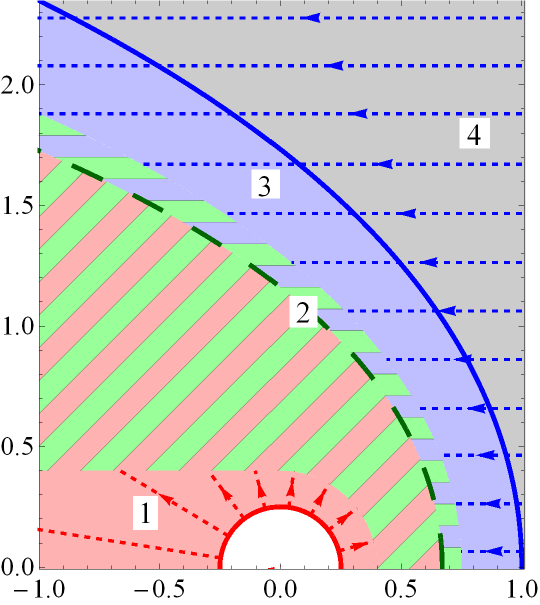}}
\caption{\label{fig:colliding-sketch}Sketch of the colliding flows region in a 3D axisymmetric Monte Carlo model of a PWN with a bow shock. Lengths are normalized to the distance from the pulsar to the bow shock apex, which is $a_0= 2.6 \times 10^{16}$ cm in the model. The white region corresponds to the cold pulsar wind zone inside the spherical termination shock of radius $R_{ts} = 0.25 a_0$ (red solid line). Red region 1 is the zone of the shocked pulsar wind (magnetic field amplitude is 20 $\mu$G, flow velocity is directed radially, $u = 0.57 c \left(R_{ts} / r_{\ast}\right)^2$, where $r_{\ast}$ is the distance from the pulsar; the flow direction is shown by red dashed lines with arrows). Gray region 4 is the unperturbed interstellar medium, blue dashed lines show the direction of the incoming flow in the rest frame of the pulsar. Blue region 3 is the interstellar medium heated by the bow shock (blue solid line; magnetic field amplitude in region 3 is 30 $\mu$G), green hatched region 2 is the zone near the contact discontinuity (magnetic field amplitude is 6 $\mu$G, the approximate position of the contact discontinuity is the green dashed line). In all regions, the magnetic field is assumed to be fully turbulent and uniform in amplitude. The mean free path of a particle $\lambda$ corresponds to the Bohm diffusion $\lambda = R_g$: in region 1 -- over the entire range of particle energies, in regions 2-3 -- up to $E_1 = 10$ GeV; at higher energies $E>E_1$ in regions 2-3 the approximation of small-scale scatterers with $\lambda = R_g E / E_1$ is assumed, where $R_g$ is the particle gyroradius. For more details on the model, see \cite{BSPWN_2017}.}
\end{figure}

Particle and antiparticle acceleration were simulated using a three-dimensional Monte Carlo model of a PWN with a bow shock \citep{BSPWN_2017,Bykov+24b,Petrov_Bykov_anis_MC_2026}. The collision zone between the pulsar's relativistic wind and the flow of ambient matter was considered as an axisymmetric system of regions with specified particle scattering parameters, turbulent magnetic field amplitudes, and flow velocity distributions (see sketch in Fig. \ref{fig:colliding-sketch}). The shape of these nested regions reproduced the outlines of a bullet-shaped region of heated pulsar wind that had passed through the termination shock and the characteristic arcuate shape of the bow shock \citep{Wilkin96}.

\begin{figure}[h!]
 \begin{minipage}{\columnwidth}
 \vspace{0pt}
   \begin{minipage}{.495\textwidth}
        \includegraphics[width=\textwidth]{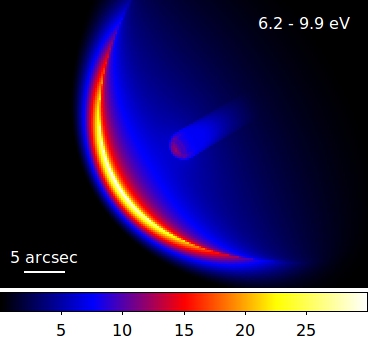}
   \end{minipage}
   \nolinebreak
   \hfill
   \begin{minipage}{.495\textwidth}
         \includegraphics[width=\textwidth]{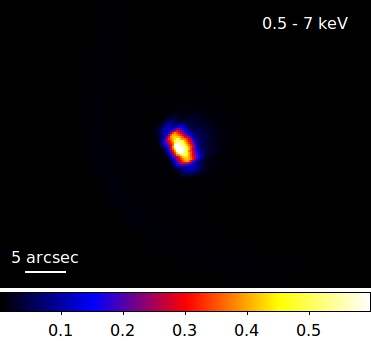}
   \end{minipage}
\vspace{0pt}
  \caption{Model maps of synchrotron emission from the BSPWN of PSR J0437-4715, built using the momentum distributions of accelerated positrons and electrons simulated throughout the nebula's volume in Monte Carlo modeling. On the left is a map of far ultraviolet emission in the 125-200 nm range; on the right is a map of X-ray emission in the 0.5-7 keV range.
  \label{fig:simages} }       
 \end{minipage}
\end{figure}

The simulation of pulsar wind's positrons and electrons acceleration in the colliding flows of the BSPWN is similar to that presented in \citep{BSPWN_2017,Bykov+19}. The non-thermal particles were injected at the front of the relativistic wind's termination shock, assuming that they were accelerated at the TS and gained the energy distribution described by the function $f_{TS}\left(E\right) = K_p E^{-2.1}$ in the range $\gamma_{min} = 4 \times 10^4 < \gamma = E / m_e c^2 < \gamma_{max} = 5 \times 10^7$. Here $K_p$ is the normalization constant, $m_e$ is the electron mass, $c$ is the speed of light. The spectral index of the particle distribution corresponds to the standard Fermi acceleration models, for which it usually lies in the range $2.0$-$2.2$. The minimum and maximum Lorentz factors of positrons and electrons in such a model are bounded below by the Lorentz factor of the cold pulsar wind $\sim 10^4$ and above by an estimate of the maximum achievable energy in accordance with the pulsar's magnetospheric potential $\sim e (\dot{E} / c)^{1/2}$, where $e$ is the positron charge (see, e.g., \citealp{Arons12}). During the Monte Carlo simulation, described in detail in Section 4 of \citep{BSPWN_2017}, each of the injected particles propagates through the collision zone of the two flows. The particle goes through a repeating cycle consisting of  (i) free motion during the mean time of free propagation and (ii) subsequent scattering (isotropic in the local rest frame of the background plasma) until it leaves the simulation region through the boundary with the free escape condition. Upon leaving the simulation region, the particle's momentum is recorded, which allows us to obtain a modeled momentum distribution of the accelerated particles produced by the object under consideration. The injection into the Fermi acceleration mechanism in the colliding flows is performed for particles with $\gamma > 2 \times 10^6$; less energetic particles are assumed to undergo advective drift into the tail of the BSPWN -- the numerical code's algorithm fixes their escape through the boundary against the axis of the simulation domain with the injection energy. Model maps of synchrotron radiation emitted in the volume of the BSPWN of PSR J0437-4715 reproduce its observed structure in the far ultraviolet and X-ray ranges (Fig. \ref{fig:simages}).

The modeling of antiproton spectrum modification differs in the implementation of the process of particle injection into the accelerator.
The antiproton spectrum measured by near-Earth observatories is assumed to be a superposition of a certain background spectrum $f_{ap}^{bkg}$ of antiprotons, unmodified in the source under consideration, and the contribution of antiprotons accelerated in the source. The background spectrum is assumed to be a power-law spectrum $f_{ap}^{bkg}\left(E\right) = K_{ap} E^{-\alpha_{1}}$, where $K_{ap}$ is a normalization constant chosen so that the flux of the model background distribution matches the observed spectrum of CR antiprotons at 50 GeV -- an energy where the solar wind modulation does not play a role. The distribution function of charged particles injected into acceleration at the front of the bow shock and accelerated by the Fermi type I mechanism can be expressed analytically:
\begin{equation}
\begin{aligned}
& f_{bow}\left(p\right) = f_{0}\left(p\right) \left(\frac{p_{inj}}{p}\right)^{\beta} + \\
& \frac{\beta}{p_{inj}} \left(\frac{p_{inj}}{p}\right)^{\beta}\int\limits_{p_{inj}}^{p}dp_0 \left(\frac{p_{0}}{p_{inj}}\right)^{\beta-1} f_{0}\left(p_0\right),
\end{aligned}
\label{eq:f-acc-bow-any-f0}
\end{equation}
where $f_0$ is the initial spectrum of particles injected from the interstellar medium, and for a typical shock $\beta = 3 u_1 / \Delta u =4$, where $u_1$ is the velocity just upstream of the shock front, $\Delta u$ is the velocity jump at the front. For a power-law spectrum $f_0 = Q_0 p^{-\alpha}$, where $Q_0$ is a normalization coefficient independent of $p$, the expression (\ref{eq:f-acc-bow-any-f0}) is simplified:
\begin{equation}
f_{bow}\left(p\right) = Q_0 p^{-\alpha}\left[1 + \frac{\alpha}{\alpha-\beta}\left(\frac{p}{p_{inj}}\right)^{\alpha-\beta}\right]
\label{eq:f-acc-bow-plaw-f0}
\end{equation}
In (\ref{eq:f-acc-bow-any-f0})-(\ref{eq:f-acc-bow-plaw-f0}) we use the distribution functions over the absolute value of momentum, which are related to the distribution function over energy as $f\left(p\right)p^2dp = f\left(E\right)dE$, so that at ultrarelativistic energies $\alpha_1 = \alpha-2$.
These expressions assume that particles are accelerated starting from a certain energy $E_{inj} = (p_{inj}^2 c^2 +m_p^2 c^4)^{1/2}$, where $m_p$ is the proton mass.
In the numerical Monte Carlo simulation of antiproton acceleration in a system of colliding flows, antiprotons are injected at the inner boundary of the layer corresponding to the vicinity of the bow shock, with an energy distribution described by (\ref{eq:f-acc-bow-plaw-f0}) and a velocity directed against the axis of the system.

Anisotropic particle diffusion in the local interstellar medium was modeled using an analytical solution of the transport equation, similar to that in \citep{Bykov+19}. The problem statement and solution of this equation, which takes into account the radiative energy losses of positrons and electrons, are given in the Appendix. The analytical consideration is possible when 
the diffusion coefficient components --
the longitudinal $D_{\parallel}$ and the transverse $D_{\perp}$ relative to the direction of the large-scale magnetic field -- have the same dependence on the particle energy, $D_{\perp}/D_{\parallel} = Const(E)$, which is consistent with the results of modeling of particle transport in chaotic magnetic fields \citep{Casse+01}. The component $D_{\parallel}$ was parameterized by Eq.(\ref{diff_meso}), the ratio $D_{\perp}/D_{\parallel} = 0.03$ \citep[see ][]{Casse+01}, and the large-scale magnetic field was assumed to be oriented in the direction $l = 52^{\circ}$, $b = 49^{\circ}$ in the Galactic coordinates, in agreement with the results of \citep{Frisch+15}. The synchrotron radiative losses of positrons and electrons were calculated assuming the amplitude of the interstellar magnetic field $B_{ism} = 3.4 \:\mu\mbox{G}$; to calculate the losses due to the inverse Compton radiation, the three-component background model used in \citep{TangPiran19} was involved, which takes into account the cosmic microwave background, the infrared and optical background radiation.

The model does not take into account the effects of heliospheric modulation of CR spectra. Therefore, when summing the contribution from the source of interest to the background antiproton spectrum, the latter is considered above energies $\sim$30 GeV, where modulation effects are insignificant. This does not affect the modeling of the contribution from PSR J0437-4715, since at energies below 30 GeV for all considered particle types it is negligible compared to the observed flux.

\section{Results}
\label{sec:results}
\begin{figure}
\center{\includegraphics[width=\columnwidth]{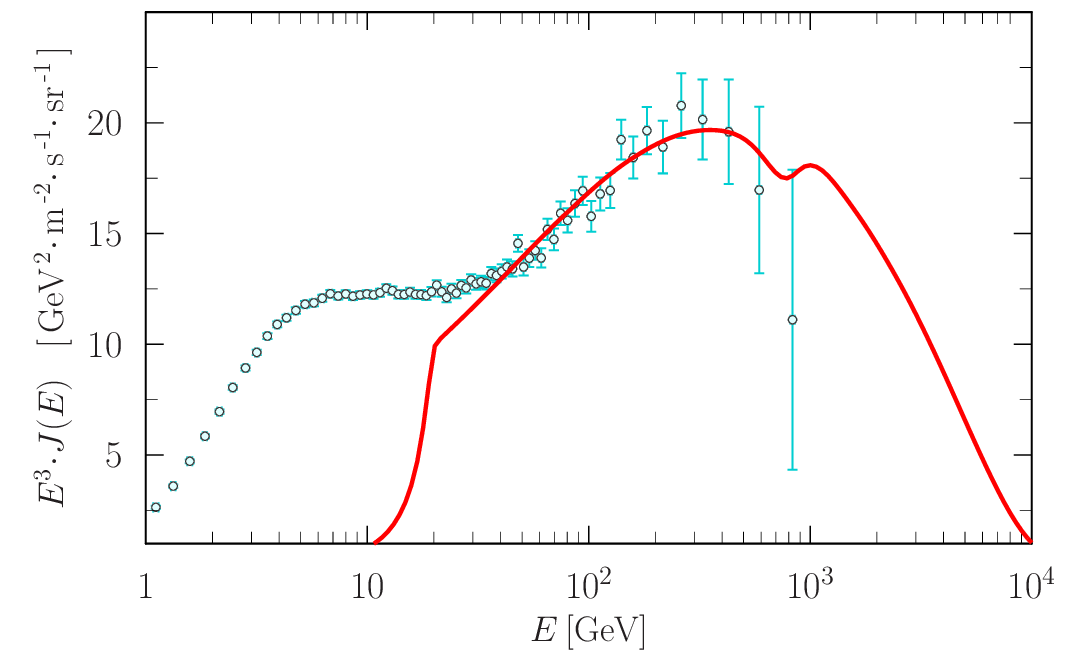}}
\caption{\label{fig:positrons}The result of modeling the near-Earth spectrum of positrons produced by the bow shock pulsar wind nebula of PSR J0437-4715 in comparison with AMS-02 data.}
\end{figure}

The results of simulation of the near-Earth positron flux with the model described above are presented in Fig. \ref{fig:positrons}. The model parameters are virtually identical to those used in \citep{Bykov+19}. The simulated positron spectrum reproduces the observed spectrum well, allowing us to attribute nearly the entire observed positron flux in the energy range from 30 GeV to 1 TeV solely to the contribution from the nearest pulsar PSR J0437-4715. According to this model, the source expends $\sim 23\% \dot{E}$ on the production of accelerated positrons and electrons.

\begin{figure}
\center{\includegraphics[width=\columnwidth]{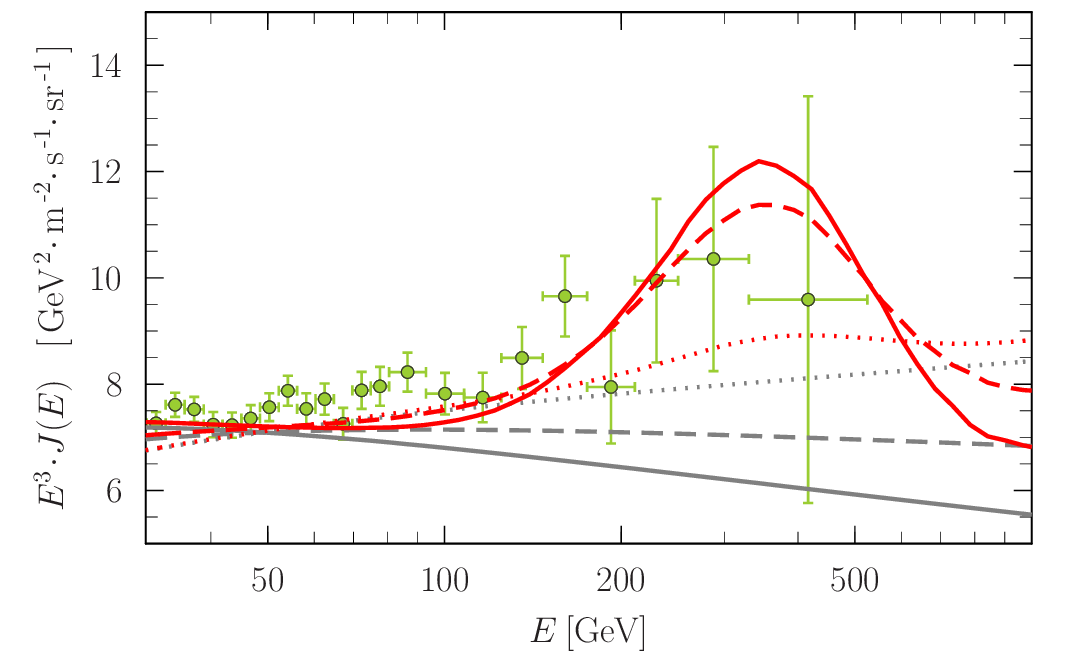}}
\caption{\label{fig:antiprotons}Results of modeling the near-Earth cosmic ray antiproton spectrum with contribution from reacceleration in the pulsar wind nebula with a bow shock of PSR J0437-4715 in comparison with AMS-02 data. Curves of different shapes present the results for three injection models, whose parameters are given in Table \ref{table:antip-injection}. Solid curves correspond to model 1 with the injection spectrum index $\alpha_1 = 3.1$, dashed curves -- to model 2 with $\alpha_1 = 3.03$, and dotted curves -- to model 3 with $\alpha_1 = 2.96$. Gray curves show the background flux of cosmic ray antiprotons that have not undergone reacceleration in the considered source (coinciding in shape with the injection spectrum), red curves show the sum of the ``background'' and the contribution from PSR J0437-4715. The models of acceleration and diffusion in the interstellar medium are completely the same as for positrons (except for the absence of radiative losses, which are negligible for hadrons).}
\end{figure}

Fig. \ref{fig:antiprotons} shows the results of a similar simulation of near-Earth antiproton fluxes including the contribution of CR antiprotons that experienced reacceleration in the colliding flows system of PSR J0437-4715. The spectra were obtained using exactly the same parameters of the Monte Carlo model of a BSPWN as for the positron spectra, i.e., using the model with the same accelerator structure and the same particle propagation parameters in the source, as well as with exactly the same model of particle diffusion in the local interstellar medium. Three different models of the injection background spectrum, unmodified by acceleration in the BSPWN of PSR J0437-4715, were considered.
The simulated antiproton fluxes allow us to reproduce the observed flux of CR antiprotons in the range of hundreds of GeV but are somewhat lower than the observed flux at tens of GeV. Less than 1\% of $\dot{E}$ is required for reacceleration of antiprotons.

\begin{table}[h!]
    \centering
    \begin{tabular}{lcccc}
    \hline       
    model & & $\alpha_1$ & & $E_{inj}$, GeV \\
    \hline
    1 & & 3.1 & & 4.5 \\
    2 & & 3.03 & & 5 \\
    3 & & 2.96 & & 12 \\
    \hline
    \end{tabular}
    \caption{Models of injection of antiprotons into the acceleration at the bow shock used in the calculations. Given are the indexes of the power law spectra of injected CR antiprotons (the background spectra, unmodified by the accelerator) and the threshold energy of injection. In all models, further injection into the acceleration in colliding flows occurs at $E_{min} \approx E_{inj}$} 
    \label{table:antip-injection}
\end{table}

\begin{figure}
\center{\includegraphics[width=\columnwidth]{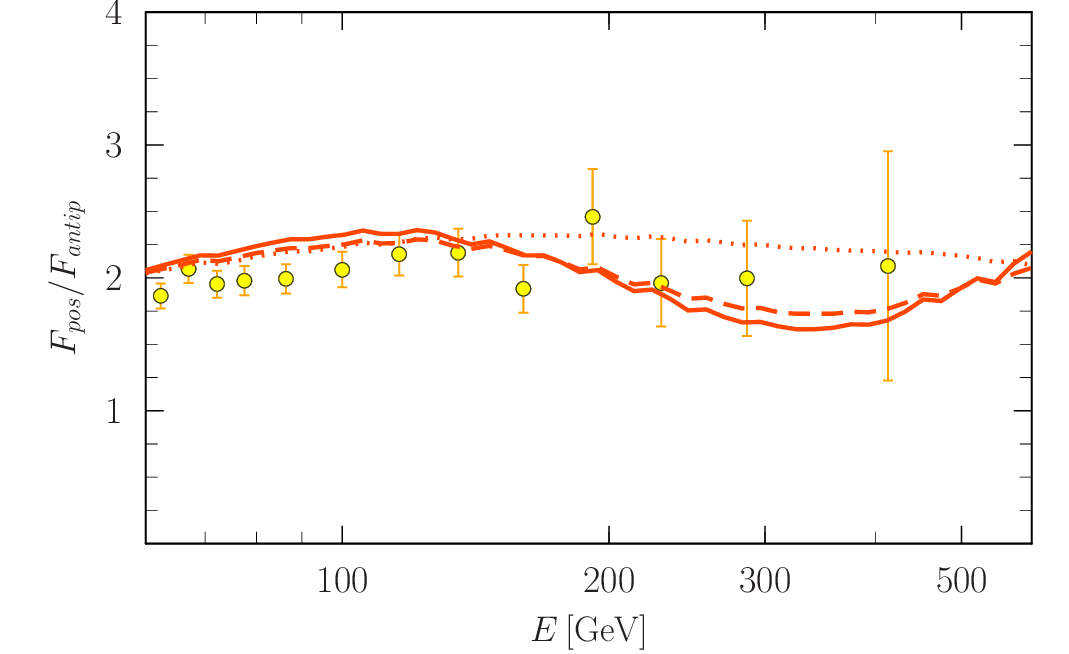}}
\caption{\label{fig:ratio-pos-to-antip}Ratio of the model near-Earth positron flux to the model near-Earth antiproton flux compared to AMS-02 data. Curves of different shapes represent the results for three injection models, the parameters of which are given in table \ref{table:antip-injection}. Solid curve correspond to model 1 with the injection spectrum index $\alpha_1 = 3.1$, dashed curve -- to model 2 with $\alpha_1 = 3.03$, and dotted curve -- to model 3 with $\alpha_1 = 2.96$.}
\end{figure}

Fig. \ref{fig:ratio-pos-to-antip} shows the model positron-to-antiproton flux ratio compared to the observed AMS-02 values for all three antiproton injection models considered. One can see that the model reproduces the observed data well in the hundreds of GeV range, while at tens of GeV, the model ratio is slightly overestimated.

\section{Discussion}
\label{sec:discussion}
Our modeling shows that, apart of producing the near-Earth GeV-TeV primary cosmic ray positrons, PSR J0437-4715 can also be a source of the near-Earth antiprotons, at least in the energy range from hundreds of GeV to TeV. The acceleration of leptons (positrons and electrons) and antiprotons requires approximately a quarter of the available power $\dot{E}$. Since both positrons and antiprotons injected into the region of colliding flows of the BSPWN gain energy in the same accelerator, the near-Earth spectral distributions of their fluxes will have a similar shape in the energy range where (i) the propagation from the source to Earth has identical character and (ii) the source contribution dominates over the background.

The propagation of positrons and electrons from the source to the Earth differs from the propagation of antiprotons by the presence of the radiative losses. For the case of PSR J0437-4715, located at a distance of 157 pc, assuming the amplitude of the interstellar magnetic field to be 3.4 $\mu$G and the diffusion coefficient $\sim 2 \times 10^{28}$ cm$^2$s$^{-1}$, one expects the characteristic time of radiative losses for leptons with energy 1 TeV to be $\sim$ 740 kyr, and their diffusion time to be less than 100 kyr. Thus, at hundreds of GeV, lepton radiative losses have little effect on the ratio of antiproton and positron fluxes.

At tens of GeV, the proposed model underpredicts the antiproton flux. This result may be due to the choice of the background antiproton flux, as well as the insufficient accuracy of the models of particles' acceleration in the source and propagation in the local interstellar medium. The applied acceleration model is phenomenological and simplified, based on a simple, prescribed by hand structure of colliding flows. The propagation model assumes uniform particle diffusion conditions throughout the pulsar's vicinity, all the way to Earth, which is also clearly an idealization. Given these simplifications, the agreement between the model spectra and the observed data appears satisfactory.
The developed model allows one to reproduce the observed positron and antiproton spectra, as well as the observed ratio of near-Earth positron and antiproton fluxes, at a level of accuracy that is reasonably consistent with the accuracy of the model itself. For a more accurate approximation of the observational data, it is necessary to provide modeling based on an accurate reproduction of the flow structure in the BSPWN of PSR J0437-4715, which is a topic for a dedicated study.

\begin{figure}
\center{\includegraphics[width=\columnwidth]{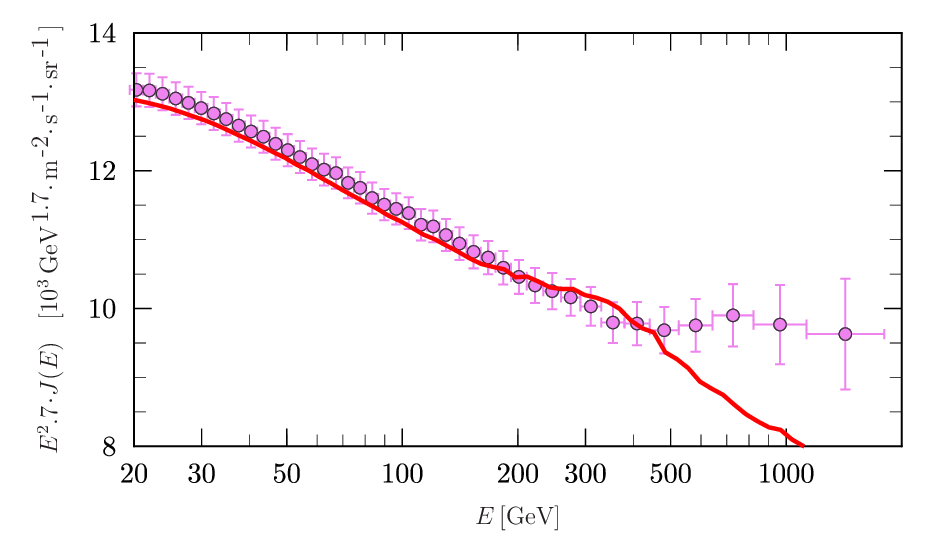}}
\caption{\label{fig:protons}
Results of modeling of the near-Earth cosmic ray proton spectrum taking into account the contribution from PSR J0437-4715. The spectrum is obtained with the discussed Monte Carlo acceleration model in the test particle approximation. The model is similar to one for the antiproton spectra and uses a power-law injection spectrum with index $\alpha_1 = 2.86$ and a normalization constant $C \approx 0.447$ m$^{-2}$sr$^{-1}$s$^{-1}$GV$^{-1}$ from approximation (12) of \citep{Aguilar+21}. Pink data points show the cosmic ray proton flux measurements by the AMS-02 observatory. The model proton flux does not contradict the detected flux, but does not explain the peculiar feature of the proton spectrum in the 0.3-30 TV range, which likely originates from a more powerful or closer source.
}
\end{figure}

\begin{figure}
\center{\includegraphics[width=\columnwidth]{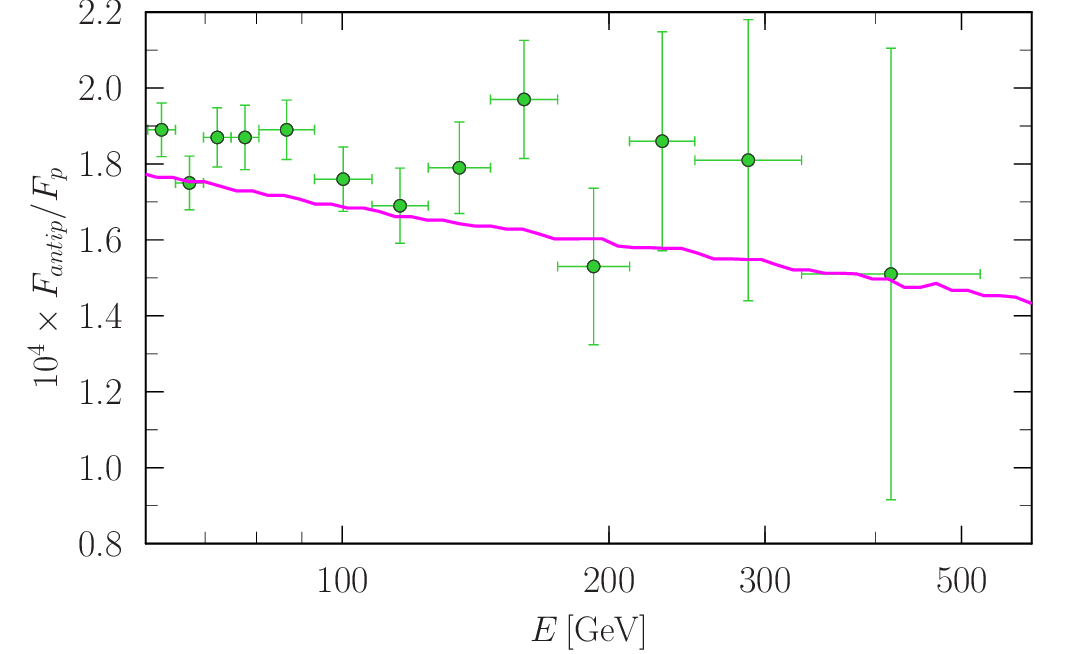}}
\caption{\label{fig:antip_to_proton}Modeled antiproton-to-proton flux ratio compared to AMS-02 data.}
\end{figure}

The observed spectra of CR protons and nuclei also contain features in the rigidity range from 0.3 to 30 TV. \citet{MalkovMoskalenko21,MalkovMoskalenko22} showed that these spectral bumps can be explained by the reacceleration of protons and nuclei at the bow shock of a nearby star located at a distance of several pc from Earth.
Simultaneously with the injection of CR antiprotons into the considered accelerator, much more numerous population of protons will be injected into it. In Fig. \ref{fig:protons}, we present the results of modeling the CR proton spectrum taking into account the contribution from  PSR J0437-4715, and in Fig. \ref{fig:antip_to_proton} we present the model antiproton-to-proton ratio. The contribution of the considered source to the CR protons was calculated using the same parameters as in the modeling of its contribution to antiprotons (injection model 3), but assuming the spectral index of the injection spectrum $\alpha_1 = 2.86$. The normalization constant was chosen to reproduce the measured amplitude of the CR proton spectrum at tens of GeV (see approximation (12) in \citealp{Aguilar+21}). The simulation showed that the proton spectrum with contribution from PSR J0437-4715 is consistent with the observed spectrum -- the perturbation produced by the considered BSPWN lies at the beginning of the teravolt bump in the observed spectrum. The production of accelerated protons in this model requires $\sim$64\%$\dot{E}$.

The sum of the powers expended on acceleration of leptons, antiprotons, and protons does not exceed the pulsar's spin-down luminosity. However, spending $\sim$87\% of the available energy budget on accelerating nonthermal particles seems to be unrealistic (especially keeping in mind the contribution from the heavier CR nuclei, which is not considered here). The model developed above is based on the approximation of test particles, which do not affect the structure of flows in the BSPWN.
The acceleration of leptons and antiprotons, which have a low energy density compared to the proton component of CR, requires a minor fraction of pulsar's spin-down power and has a relatively weak effect on the accelerator structure. Protons, being the energetically dominant component, can significantly influence on the accelerator. 
A nonlinear kinetic model of particle acceleration in the collision zone of two converging plane-parallel shocks, taking into account the feedback of the accelerated protons on the flow structure \citep{Bykov+13}, demonstrates the possibility of a highly efficient conversion of the accelerator's power (significantly greater than 10\%) into relativistic particles. In the case of a bow shock, the flow geometry in the accelerator is more complex. A more realistic model of proton (and heavy nuclei) acceleration should simultaneously and consistently calculate the structure of the BSPWN, taking into account the pressure and currents of nonthermal particles and the momentum distributions of these particles.

The test particle simulations with reasonable parameters of the acceleration model in the BSPWN showed that one would need a power comparable with $\dot{E}$ of PSR J0437-4715 to explain {\it the entire} teravolt feature in the range from 0.3 to 30 TV in the spectra of CR protons and nuclei by this pulsar's contribution. Accounting for the nonlinear effects modifies the spectra of accelerated particles and the flow structure.
On the other hand, as suggested by \citet{MalkovMoskalenko21,MalkovMoskalenko22}, some of accelerators located closer to the Earth, such as the star Epsilon Eridani at a distance of about 3 pc, may be responsible for the features of the CR proton spectrum in the TeV range. Such stars, however, are not the sources of primary CR positrons, unlike the nearby pulsar PSR J0437-4715.

Estimates of the characteristic timescales of particles' diffusion and radiative losses, similar to those given above, allow us to conclude that only pulsars located at distances no more than 400-500 pc from Earth may be sources of the near-Earth CR positrons with an energy of $\sim$ 1 TeV. In addition to PSR J0437-4715, other possible candidates for the sources of detected GeV-TeV cosmic ray antiparticles include the Geminga pulsar, PSR B0656+14, PSR J1741-2054, and several others. The young Vela pulsar has a significantly higher spin-down power and can likely also be classified as a source of the type considered in this paper, since its nebula is surrounded by a high-speed plasma flow inside the parent supernova remnant. Taking into account the existing estimates of the age of the Vela pulsar and the possible effect of the expanding supernova remnant on the flux of antiprotons injected into acceleration, the role of this powerful pulsar as a source of the observed antiparticles will be studied separately.

\section{Conclusions}\label{sec:conclusion}
Modeling of the contribution of the bow shock PWN of the nearest pulsar PSR J0437-4715 to the observed near-Earth fluxes of CR antiparticles -- positrons and antiprotons -- demonstrates that the nearly energy-independent positron-to-antiproton flux ratio can be explained by models involving particle acceleration in a BSPWN. Such models should be considered alongside theories of the origin of primary antiparticles in the dark matter annihilations/decays. A BSPWN can not only accelerate positrons and electrons from the pulsar wind, but also re-accelerate CR particles produced in other sources. These acceleration processes occur in the same BSPWN flow system, resulting in identical spectral shapes of accelerated particles and a nearly energy-independent ratio of the fluxes up to TeV energies, at which the radiative energy losses of leptons come into play.

PSR J0437-4715 is a natural source of primary positrons, being the closest known pulsar to Earth. Calculations with reasonable model parameters of particle diffusion in the local interstellar medium have shown that contribution of PSR J0437-4715 can explain the observed CR positron flux in the range from 30 to 1000 GeV. Moreover, the production of accelerated leptons (both positrons and electrons) would consume no more than a quarter of the pulsar's spin-down power. This source can simultaneously accelerate CR antiprotons, ensuring a weak energy dependence of the flux ratio between these two types of antiparticles. Less than 1\% of the available power is expended on the production of accelerated antiproton spectrum.

\section*{Acknowledgements}
Modeling of the spectra of near-Earth antiprotons and positrons by A.E.P. was supported by the Foundation for the Advancement of Theoretical Physics and Mathematics ``BASIS'' (grant no. 24-1-3-28-1). A model of the plasma flow structure in the BSPWN of PSR J0437-4715 was developed by A.M.B. supported by the RSF grant 25-72-20007. Numerical modeling was partially performed using the resources of the supercomputer center of St. Petersburg Polytechnic University http://scc.spbstu.ru. The authors thank M.E. Kalyashova for helpful comments on a draft of the article.

\appendix

\section{Appendix} \label{sec:appendix}
Let us consider the diffusion of particles from a source located at the origin of a cylindrical coordinate system with the Oz axis directed along the local large-scale interstellar magnetic field. The particle transport equation is:
\begin{equation}
\begin{aligned}
\frac{\partial F}{\partial t} = &\frac{1}{r}\frac{\partial}{\partial r} \left(r D_{\perp}\left(p\right) \frac{\partial F}{\partial r}\right) + \frac{\partial}{\partial z} \left(D_{\parallel}\left(p\right) \frac{\partial F}{\partial z}\right) \\ &+ \frac{1}{p^2} \frac{\partial}{\partial p}\left(p^2 b\left(p\right) F\right) + Q(\textbf{r},p,t).
\end{aligned}
\label{initial_transport_equation}
\end{equation}
where $F\left(\textbf{r},p\right)$ is the particle distribution function with the particles' number density $n_0 = \int F d^3p$; $t$, $\textbf{r}$, $p$ are the time, radius vector of the point in space, and particle momentum, respectively; $r$, $z$ are the cylindrical components of $\textbf{r}$; $D_{\parallel}$, $D_{\perp}$ are the components of the diffusion coefficient parallel and perpendicular to the large-scale magnetic field; $b \left(p\right)$ is the radiative energy loss rate of particles, and the function $Q$ defines the flux density of accelerated particles injected into the system by the source. In the case of a stationary source located at the origin, it can be written as
\be
Q\left(\textbf{r}, p\right) = J\left(p\right)\delta\left(\textbf{r}\right),
\label{source}
\ee
where $\delta$ is the Dirac's delta-function.

For simplicity, and also based on the results of Monte Carlo simulations of CR particle transport in chaotic magnetic fields \citep{Casse+01}, we assume the same dependence of the diffusion coefficients on the momentum, introducing the dimensionless function $g\left(p\right) = D_{\parallel} / D_{\parallel, 0} = D_{\perp} / D_{\perp, 0}$.
We normalize (\ref{initial_transport_equation}) using an arbitrary time scale $T_0$ and spatial scales $R_0 = \sqrt{D_{\perp,0} T_0}$, $Z_0 = \sqrt{D_{\parallel,0} T_0}$, introducing the variables $\tau = t / T_0$, $\rho = r / R_0$, $\zeta = z / Z_0$:
\be
\hspace{-2mm}\frac{\partial f}{\partial \tau} \hspace{-0.5mm} = \hspace{-0.5mm} \tilde{g}\left(\gamma\right) \Delta f \hspace{-0.5mm} -  \frac{\partial}{\partial \gamma}\hspace{-1mm} \left(\tilde{b}\left(\gamma\right) f\right) \hspace{-0.2mm} + \hspace{-0.2mm} \frac{p^2 J\left(p\right)\delta\left(\rho\right)\delta\left(\zeta\right)}{2\pi R_0^2 Z_0 \rho c / T_0},
\label{transport_equation_normalized}
\ee
where $c$ is the speed of light, $\Delta = \rho^{-1}\partial_{\rho}\left(\rho\partial_{\rho}\right) + \partial_{\zeta}^2$ is the Laplace operator, $\gamma = p / mc$ is the Lorentz factor of the particle, $m$ is the mass of the particle, and the functions $f\left(\gamma\right) = F\left(p\right)p^2 / c$, $\tilde{b}\left(\gamma\right) = - T_0 \, b\left(m c \gamma\right) / m c$, $\tilde{g}\left(\gamma\right) = g\left(m c \gamma\right)$.

The Green's function of the transport equation (\ref{transport_equation_normalized}), given by the formula (\ref{Green_function_full}), was obtained by \citet{GinzburgSyrovatskii64} (see also \citealp{Berezinskii_astrophys_of_CR_1990}):
\be
\ba
& \hspace{-3mm} G\left(\gamma, \rho, \zeta, \tau; \, \gamma_0, \rho_0, \zeta_0, \tau_0\right) = \\ & \hspace{-3mm}
\frac{\delta\left(\tau - \tau_0 - \chi\right)}{\left(4\pi \lambda\left(\gamma,\gamma_0\right)\right)^{3/2}\hspace{-0.3mm}\left|\tilde{b}\left(\gamma\right)\right|}
\hspace{-0.3mm}\exp\hspace{-0.3mm} \left[\hspace{-0.1mm} - \frac{\left(\rho-\rho_0\right)^2 + \left(\zeta - \zeta_0\right)^2}{4\lambda\left(\gamma,\gamma_0\right)}\right]
\label{Green_function_full}
\ea
\ee
where the functions $\chi$ and $\lambda$ are defined as:
\be
\chi\left(\gamma,\gamma_0\right) = \int\limits_{\gamma_0}^{\gamma}\frac{d \gamma'}{\tilde{b}\left(\gamma'\right)};
\label{chi_function}
\ee
\be
\lambda\left(\gamma,\gamma_0\right) = \int\limits_{\gamma_0}^{\gamma}\frac{\tilde{g}\left(\gamma'\right)}{\tilde{b}\left(\gamma'\right)}d \gamma'.
\label{lambda_function}
\ee

By integrating (\ref{Green_function_full}) over $\tau_0$, one can obtain the Green's function (\ref{Green_function_stationary}) for the stationary problem.
\be
\ba
& \hspace{-2.5mm} G\left(\gamma, \rho, \zeta; \, \gamma_0, \rho_0, \zeta_0\right) = \\ & \hspace{-2.5mm} \left\{
\begin{aligned}
&\frac{\exp\hspace{-1mm}\left[\hspace{-0.1mm} - \hspace{-0.5mm}\left(\left(\rho-\rho_0\right)^2 \hspace{-0.5mm}+\hspace{-0.5mm} \left(\zeta - \zeta_0\right)^2\right)/4\lambda\left(\gamma,\gamma_0\right)\right]}{\left(4\pi \lambda\left(\gamma,\gamma_0\right)\right)^{3/2}\left| \tilde{b}\left(\gamma\right) \right|},\chi > 0 \\
& 0, \chi \leq 0
\end{aligned}
\right.
\ea
\label{Green_function_stationary}
\ee

To obtain the distribution function $f$ from (\ref{Green_function_stationary}), it is necessary to calculate the integrals given in the equation (\ref{f_distr_integrals_G_source}), where $Q_{norm}$ is the last term in (\ref{transport_equation_normalized}), $p_0 = m c \gamma_0$:
\be
\ba
& f\left(\gamma, \rho, \zeta\right) = 2 \pi \int\limits_{-\infty}^{+\infty} d\zeta_0 \int_{0}^{+\infty} \hspace{-1mm} \rho_0 d\rho_0 \int_{0}^{+\infty} d \gamma_0 \times \\ & G\left(\gamma, \rho, \zeta; \, \gamma_0, \rho_0, \zeta_0\right) Q_{norm}\left(\rho_0,\zeta_0,p_0\right)
\label{f_distr_integrals_G_source}
\ea
\ee
After simple integration we can obtain the expression (\ref{f_distr}) for $f$, where $\Theta$ is the Heaviside function.
\be
\ba
& f\left(\gamma, \rho, \zeta\right) =  \frac{1}{c D_{\perp,0}\sqrt{D_{\parallel,0} T_0}}
\int\limits_{0}^{+\infty} d \gamma_0 \cdot p_0^2 J\left(p_0\right) \cdot \\ & \frac{\Theta\left[\chi\left(\gamma,\gamma_0\right)\right]}{\left(4\pi\lambda\left(\gamma,\gamma_0\right)\right)^{3/2}\left|\tilde{b}\left(\gamma\right)\right|}
\cdot \exp\left[- \frac{\rho^2+\zeta^2}{4\lambda\left(\gamma,\gamma_0\right)}\right]
\label{f_distr}
\ea
\ee

For antiprotons and nuclei the radiative losses are negligible, so the equation (\ref{initial_transport_equation}) for a stationary source after a similar spatial scaling with $R_0 = \sqrt{D_{\perp} T_0}$, $Z_0 = \sqrt{D_{\parallel} T_0}$ simplifies to the Poisson equation, the solution of which is
\be
F\left(\textbf{r},p\right) = \frac{J\left(p\right)}{4\pi D_{\perp} \sqrt{z^2 + \left(D_{\parallel} / D_{\perp}\right) r^2}}.
\label{f_distr_protons}
\ee

%

\end{document}